\documentclass[preprint]{revtex4}
\usepackage{graphicx}
\usepackage{subfigure}
\usepackage{epsfig}
\usepackage{bigints}
\usepackage[usenames]{color}

\begin{document}

\title{Cavitation in a quark gluon plasma with finite chemical potential and several transport coefficients}

\author{S. M. Sanches Jr.\dag\,, D. A. Foga\c{c}a\dag\,, F. S. Navarra\dag\ and H. Marrochio\dag }
\address{\dag\ Instituto de F\'{\i}sica, Universidade de S\~{a}o Paulo\\
Rua do Mat\~ao Travessa R, 187, 05508-090, S\~{a}o Paulo, SP, Brazil}

\begin{abstract}

We study the effects of a finite chemical potential on the occurrence of cavitation in a quark gluon plasma  (QGP).
We solve the evolution equations of second order viscous relativistic hydrodynamics using two equations of state.  The first one was derived in  lattice QCD and represents QGP at zero chemical potential and it was previously
used in  the  study of cavitation. The second equation of state also comes from lattice QCD and is a recent parametrization of
the QGP at finite chemical potential. We conclude that at finite chemical potential  cavitation in the QGP
occurs earlier  than at zero chemical potential.  We also consider transport coefficients from a holographic model of a non-conformal
QGP at zero chemical potential.  In this case cavitation does not occur.

\end{abstract}

%\pacs{PACS Numbers : 12.38.Mh, 12.39.Ki, 25.75.-q }

\maketitle

\section{Introduction}

The analysis of an extensive body of experimental information suggests that in the heavy ion collisions
at high energies performed at the Relativistic Heavy Ion Collider (RHIC) \cite{rhic1,rhic2} and at the Large Hadron Collider
(LHC) \cite{lhc} a quark gluon plasma (QGP) is formed and it behaves like an almost perfect fluid \cite{schu}, with small dissipation.
This highly energetic fluid is very well described by relativistic hydrodynamics \cite{roma1,heinz,scha,gale}.
Some years ago it has been pointed out \cite{torri1,torri2,raja} that previously overlooked viscous effects could lead to processes such as
cavitation in these systems.

In general it is well known \cite{livrocav} that if a non-relativistic liquid at a constant pressure reaches the critical temperature, the consequent
ruptures of  cohesion between molecules leads to boiling.  On the other hand, cavitation occurs when at a constant temperature the critical pressure
drops below the vapor pressure and vapor bubbles are formed.

In an expanding quark gluon plasma, because of a non-vanishing viscosity, the pressure can become zero and even negative.  In this situation cavitation
would imply the creation of bubbles of a hadron gas phase in the quark gluon plasma phase. The evolution of these bubbles might have observable effects
\cite{romacav}.

The study of cavitation in the QGP was pioneered by the authors of Refs. \cite{torri1,torri2} and further developed in Ref. \cite{raja}.
Bulk viscosity has been a very active topic of research in heavy ion physics \cite{bulkgeral}. It is generally understood that for QCD at high
temperatures, bulk viscosity effects are  much smaller than shear ones, as for high temperatures the conformal limit is well approximated
\cite{bulkgeral}.  However, near the deconfinement region and in hadronic systems, large values of the bulk contribution have been found in
several works \cite{bulkhad}.  These large values might trigger cavitation in the evolving plasma.

In the present work we extend the formalism employed in \cite{raja} to the case of finite chemical potential. From phenomenological
studies of particle production in heavy ion collisions with the help of thermal models \cite{andro1} we can conclude that the baryon chemical
potential is of the order of $\mu_B \simeq 1-10$ MeV at LHC and $\mu_B \simeq 20-100$ MeV at RHIC. At lower energies, the chemical
potential can reach higher values, $\mu_B \simeq 500$ MeV. Moreover, in the case of inhomogeneous freeze-out, even at high energies we can
have domains of high chemical potential \cite{licinio}.  Recently the equation of state (EOS) of the QGP at finite chemical potential  was
computed in lattice QCD simulations \cite{ratti,fodor14}. In view of these results, we think that it is interesting to investigate the
effect of finite chemical potential on cavitation.  In the search for different phenomenological parametrizations of the bulk viscosity coefficient
and other transport parameters, based on different many-body microscopic dynamics, we have considered the holographic non-conformal model
of QGP of Ref. \cite{hugo14}.

\section{Second order relativistic viscous hydrodynamics}

\subsection{The main equations}

Throughout this study we use natural units $\hbar=c=k_{B}=1$. The energy momentum tensor of  viscous relativistic hydrodynamics
\cite{raja,deni15,romacav,nlcw,hugo14}  is:
\begin{equation}
T^{\mu \nu}=\varepsilon u^\mu u^\nu - p\Delta^{\mu \nu} + \Pi^{\mu \nu}
\label{EMTensor}
\end{equation}
where the $\varepsilon$ and $p$ are the energy density and the pressure of the fluid, $u^{\mu}=(\gamma,\gamma\vec{v})$ is the
4-velocity of the fluid ($\gamma$ is the Lorentz factor $\gamma=(1-v^{2})^{-1/2}$) with normalization $u_{\mu}u^{\mu}=-1$. The metric is given by
$g_{\mu\nu}=\textrm{diag}(-,+,+,+)$.
$\Pi^{\mu \nu}$ is the viscous tensor satisfying $u_{\mu} \Pi^{\mu \nu} = 0$ and $\Delta^{\mu \nu} \equiv g^{\mu \nu} + u^{\mu} u^{\nu}$
is the  projector orthogonal to $u^{\mu}$.  Notice that we do not include dissipative currents associated with finite chemical potential and
heat flow. In Bjorken flow \cite{bjk},  symmetry constraints eliminate these contributions from the equations of motion \cite{muronga03}. The heat flux
$q^{\mu}$ satisfies the orthogonality property $q^{\mu}u_{\mu}=0$ in the Landau-Lifshitz frame \cite{muronga03}.  The two relevant terms in the coupling
of the dissipative equations involve the terms $\nabla_{< \, \mu}q_{\nu \, >}$ and $\nabla^{\perp}_{\mu}q^{\mu}$.  In the Bjorken flow the temporal component
of the heat flux is zero ($q^{\tau}=0$), which leads us to the conclusion that the previous heat flux contributions are exactly zero \cite{muronga03}.

Conservation of the energy and momentum is enforced by $\nabla_{\mu}T^{\mu \nu}=0$, where $\nabla_{\mu}$ is the covariant derivative.
 The projection of this equation on the direction parallel
to the fluid velocity reads: $u_{\nu}\nabla_{\mu}T^{\mu \nu}=0$. Similarly,  the projection on the direction
perpendicular  to the fluid velocity yields: $\Delta_{\nu}^{\alpha}\nabla_{\mu}T^{\mu \nu}=0$. These two projections lead to
the following evolution equations \cite{raja,nlcw,hugo14}:
\begin{equation}
D\varepsilon+(\varepsilon+p)\theta+\Pi^{\mu \nu}\nabla_{\perp \, (\mu} u_{\nu)}=0
\label{evoeq1}
\end{equation}
and
\begin{equation}
(\varepsilon+p)Du^{\mu}+\nabla_{\perp}^{\mu}p+\Delta^{\mu}_{\alpha}\nabla_{\beta}\Pi^{\alpha\beta}=0  \,\,\, .
\label{evoeq2}
\end{equation}
In the above equations we have $D\equiv u^{\mu}\nabla_{\mu}$, $\theta \equiv \nabla_{\mu}u^{\mu}$,
$\nabla_{\perp}^{\mu} \equiv \Delta^{\mu \alpha}\nabla_{\alpha}$ and the symmetrization of the indices is represented by
$A_{(\mu \nu)} = (A_{\mu \nu} + A_{\nu \mu})/2$.

The viscous tensor $\Pi^{\mu \nu}$ is given by \cite{raja,romacav,hugo14}:
\begin{equation}
\Pi^{\mu\nu}=\pi^{\mu \nu}+\Pi\Delta^{\mu \nu}
\label{previscotens}
\end{equation}
where $\pi^{\mu \nu}$ is the traceless symmetric part $\pi^{\nu}_{\nu}=0$ and $\Pi\equiv g_{\mu \nu} \Pi^{\mu \nu}/3$.
We obtain the components of the viscous tensor (\ref{previscotens})  solving the following equations \cite{raja,nlcw,hugo14} which are
second order Israel-Stewart like \cite{ist}:
\begin{equation}
\tau^{\eta}_{\Pi}\bigg(D \pi^{\langle \mu \nu \rangle}+{\frac{4}{3}}\theta \pi^{\mu \nu} \bigg)+\pi^{\mu \nu}=-2\eta \sigma^{\mu \nu}
+{\frac{\lambda_{1}}{\eta^{2}}}\pi_{\lambda}^{\langle \mu} \pi^{\nu \rangle \lambda}
\label{viscotens1}
\end{equation}
\begin{equation}
\Pi=-\zeta \theta -\tau^{\zeta}_{\Pi}D\Pi
\label{viscotens2}
\end{equation}
with the definitions $\sigma^{\mu \nu}\equiv \Delta^{\mu \nu \alpha \beta}\nabla_{\alpha}u_{\beta}$, \,
$\Delta^{\mu \nu \alpha \beta}\equiv \big( \Delta^{\mu \alpha}\Delta^{\nu \beta}+
\Delta^{\mu \beta}\Delta^{\nu \alpha}\big)/2-\Delta^{\mu \nu} \Delta^{\alpha \beta}/3$ \, and
$A^{\langle \mu \nu \rangle} \equiv \Delta^{\mu \nu \alpha \beta} A_{\alpha \beta} \,\,\, .$

The shear viscosity is $\eta$, and the second-order coefficients are
$\tau^{\eta}_{\Pi}$ and $\lambda_{1}$, given by the holographic calculations of $\mathcal{N}=4$ SYM \cite{kss,brsss}:
\begin{equation}
{\frac{\eta}{s}}={\frac{1}{4\pi}}
\label{etapers}
\end{equation}
\begin{equation}
\tau^{\eta}_{\Pi}={\frac{[2-ln(2)]}{2\pi T}}
\label{taueta}
\end{equation}
\begin{equation}
\lambda_{1}={\frac{\eta}{2\pi T}}={\frac{s}{8\pi^{2}T}}
\label{lambdaum}
\end{equation}
and by the following parametrization of $\zeta/s$ \cite{raja,deni15}:
\begin{equation}
{\frac{\zeta}{s}}=a\, exp\Bigg( {\frac{T_{c}-T}{\Delta T}} \Bigg)+
b\Bigg({\frac{T_{c}}{T}}\Bigg)^{2} \,\,\,\,\,  \textrm{for}  \,\,\, T>T_{c}
\label{zetapers}
\end{equation}
whith $a=0.901$, $b=0.061$ and $\Delta T=T_{c}/14.5$ \, .  We also consider $\tau^{\eta}_{\Pi}=\tau^{\zeta}_{\Pi}$.
We obtain the entropy density  from the relation
\begin{equation}
s={\frac{\partial p}{\partial T}} \,\,\, .
\label{entd}
\end{equation}
For this parametrization we will choose $T_{c}=190 \sim 200 \, MeV$.
Setting the second-order coefficients equal to zero ($\tau^{\eta}_{\Pi}=0$, $\tau^{\zeta}_{\Pi}=0$ and $\lambda_{1}=0$) we recover
the Navier-Stokes approach in (\ref{viscotens1}) and (\ref{viscotens2}): $\pi^{\mu \nu}=-2\eta\sigma^{\mu \nu}$ and $\Pi=-\zeta \theta$.

\subsection{Bjorken flow}

Now we consider  solutions of the hydrodynamical equations which do not depend on the transverse spatial coordinates $x$ and $y$
and are boost invariant in the $z-$direction as in \cite{raja,deni15,hugo14}, which are the Bjorken symmetries. To obtain the evolution
equations in this boost invariant description we perform the changes of variables from $(t,z)$ to $(\tau,\xi)$, where $\tau$ is the
proper time and $\xi$ is the spacetime rapidity, given by \cite{bjk}: $\tau \equiv \sqrt{t^{2}-z^{2}}$ and
$\xi \equiv \textrm{tanh}^{-1}(z/t)$.  In general, the Milne coordinates are $(\tau,r,\phi,\xi)$ and the line element is
$ds^{2} = -d\tau^{2} + dr^{2} + r^{2}d\phi^{2} + \tau^{2}d\xi^{2}$ so $g_{\mu \nu}=diag(-1,1,1,\tau^{2})$. Due to the Bjorken symmetry
the four-velocity in Milne coordinates turns out to be $u^{\mu} = (1, 0, 0, 0)$.

In the $(\tau,\xi)$ coordinates we have $\theta=1/\tau$, $D=\nabla_{\tau}$, $\Gamma^{\xi}_{\tau \xi}=\Gamma^{\xi}_{\xi \tau}=1/\tau$,
$\Gamma^{\tau}_{\xi \xi}=\tau$ and $\nabla_{\mu} u_{\nu}=-\Gamma^{\lambda}_{\mu\nu}u_{\lambda}=\Gamma^{\tau}_{\mu \nu}$ .
In particular,  $\nabla_{\xi} u_{\xi}=\tau$ .

The energy-momentum tensor (\ref{EMTensor}) becomes:
\begin{equation}
 T^{\mu}_{\nu}=
\left( \begin{array}{cccc}
\varepsilon & 0 & 0 & 0 \\
0 & p & 0 & 0 \\
0 & 0 & p & 0 \\
0 & 0 & 0 & p \\ \end{array} \right) +
\left( \begin{array}{cccc}
0 & 0 & 0 & 0 \\
0 & \Pi + \frac{\Phi}{2} & 0 & 0 \\
0 & 0 & \Pi + \frac{\Phi}{2} & 0 \\
0 & 0 & 0 & \Pi - \Phi \\ \end{array}  \right)
\label{tmunibjk}
\end{equation}
where the contributions to the pressure from the bulk and shear stresses are respectively given by $\Pi$ (the trace of $\Pi^{\alpha\beta}$)
and $\Phi$ (the traceless part of $\Pi^{\alpha\beta}$), given by $\pi^{x}_{x}=\pi^{y}_{y}\equiv \Phi/2$ and $\pi^{\xi}_{\xi}=-\Phi$.

From (\ref{tmunibjk}) we identify the transverse pressure:
\begin{equation}
P_{\perp} \equiv p + \Pi + \frac{\Phi}{2}
\label{ptrans}
\end{equation}
and the longitudinal pressure:
\begin{equation}
P_{\xi} \equiv p + \Pi - \Phi
\label{plon}
\end{equation}

In the $(\tau,\xi)$ description, the equations (\ref{evoeq1}), (\ref{evoeq2})
 and (\ref{previscotens}) give the following evolution equations \cite{raja}:
\begin{equation}
\frac{\partial \varepsilon}{\partial \tau} = - \frac{\varepsilon + p + \Pi - \Phi}{\tau}
\label{evoener}
\end{equation}
\begin{equation}
\tau^{\eta}_{\Pi} \frac{\partial \Phi}{\partial \tau} = \frac{4 \eta}{3 \tau} - \Phi - \left[ \frac{4 \tau^{\eta}_{\Pi}}{3 \tau} \Phi
+ \frac{\lambda_{1}}{2 \eta^2} \Phi^2 \right]
\label{evophi}
\end{equation}
and
\begin{equation}
\tau^{\zeta}_{\Pi} \frac{\partial \Pi}{\partial \tau} = - \frac{\zeta}{\tau} - \Pi \,\,\, .
\label{evopi}
\end{equation}
Cavitation will take place when the longitudinal pressure $P_{\xi}$ becomes negative.

\section{Equation of state}

In this work we study the occurrence of cavitation considering two equations of state for the quark gluon plasma.

\subsection{Model 1}

The following EOS was previously employed in \cite{raja} and we use it to perform comparison with other models.  This equation of state
describes the quark gluon plasma phase and the crossover to a hadron gas, being a  parametrization of a lattice calculation  at zero
chemical potential with $T_c=190$ $MeV$.  The trace anomaly is given by \cite{baza}:
\begin{equation}
{\frac{\varepsilon_{1}-3p_{1}}{T^{4}}}= \left(1-\frac{1}
{\left[1+\exp\left(\frac{T-c_1}{c_2}\right)\right]^2}\right)
\left(\frac{d_2}{T^2}+\frac{d_4}{T^4}\right)\
\label{rajaTraceAnomaly}
\end{equation}
$d_2= 0.24$ $GeV^2$, $d_4=0.0054$ $GeV^4$, $c_1=0.2073$ $GeV$ and $c_2=0.0172$ $GeV$.  The pressure  \cite{raja} is:
\begin{equation}
p_{1}= T^{4} \, \bigintss_{T_0}^T dT' \,\,
\left(1-\frac{1}
{\left[1+\exp\left(\frac{T^{'}-c_1}{c_2}\right)\right]^2}\right)
\left(\frac{d_2}{T^{' \, 3}}+\frac{d_4}{T^{' \, 5}}\right)
\label{rajapre}
\end{equation}
From (\ref{rajaTraceAnomaly}) and (\ref{rajapre}) we obtain the energy density:
$$
\varepsilon_{1}=T^{4} \left(1-\frac{1}
{\left[1+\exp\left(\frac{T-c_1}{c_2}\right)\right]^2}\right)
\left(\frac{d_2}{T^2}+\frac{d_4}{T^4}\right)
$$
\begin{equation}
+ \, 3 \, T^{4} \, \bigintss_{T_0}^T dT' \,\,
\left(1-\frac{1}
{\left[1+\exp\left(\frac{T^{'}-c_1}{c_2}\right)\right]^2}\right)
\left(\frac{d_2}{T^{' \, 3}}+\frac{d_4}{T^{' \, 5}}\right)
\label{rajaeps}
\end{equation}

\subsection{Model 2}

In this work we study finite chemical potential effects in cavitation.  To do so, we use the recent parametrization of a lattice simulation of
$SU(3)$ QCD matter at finite temperature and chemical potential, with  quarks ($u$, $d$ and $s$ with equal masses) and gluons \cite{ratti,fodor14}.

We have previously used this equation of state in the study of the expansion of the early Universe in \cite{we2015}, where we have developed the
 expressions in detail. We start by recalling  the trace anomaly at finite chemical potential \cite{ratti,fodor14}:
$$
{\frac{{\varepsilon}_{2}(T,\mu)-3\, {p}_{2}(T,\mu)}{T^{4}}}=T{\frac{\partial}{\partial T}}
\Bigg[{\frac{{p}_{2}(T,\mu)}{T^{4}}}\Bigg]+{\frac{\mu^{2}}{T^{2}}}\,\chi_{2}
$$
\begin{equation}
={\frac{{\varepsilon}_{2}(T,0)-3\, {p}_{2}(T,0)}{T^{4}}}
+{\frac{\mu^{2}}{2T}}\,{\frac{d \chi_{2}}{d T}}
\label{fodortrami}
\end{equation}
where the chemical potential contribution is given by the function \cite{ratti}:
\begin{equation}
\chi_{2} (T)= e^{-h_{3}/{\tau}-h_{4}/{\tau}^{2}} \cdot f_{3} \cdot \Big[tanh (f_{4}\cdot {\tau}+f_{5})+1\Big]
\label{chemit}
\end{equation}
In the zero chemical potential limit we have \cite{ratti,fodor14}:
\begin{equation}
{\frac{{\varepsilon}_{2}(T,0)-3\, {p}_{2}(T,0)}{T^{4}}}=e^{-h_{1}/{\tau}-h_{2}/{\tau}^{2}} \cdot \Bigg[h_{0}+
{\frac{f_{0} \cdot \Big[tanh(f_{1}\cdot {\tau}+f_{2})+1\Big]}{1+g_{1} \cdot {\tau}+g_{2} \cdot {\tau}^{2}}}   \Bigg] \,\,\, .
\label{fodortrazero}
\end{equation}
In the last three expressions we have introduced the variable $\tau=T/200 $ $MeV$, where $200 $ $MeV$ is the critical temperature.

We consider two sets for the dimensionless parameters:

{\it{(i)}} From \cite{fodor14}: $h_{0} = 0.1396$, $h_{1} = -0.1800$, $h_{2} = 0.0350$, $f_{0} = 1.05$,
$f_{1} = 6.39$, $f_{2} = -4.72$, $g_{1} = -0.92$ and
$g_{2} = 0.57$ . For the chemical potential parametrization we use from \cite{ratti}: $h_{3} = -0.5022$, $h_{4} = 0.5950$, $f_{3} = 0.1359$, $f_{4} = 6.3290$ and $f_{5} = -4.8303$ .

{\it{(ii)}} From \cite{ratti}:
 $h_{0} = 0.1396$, $h_{1} = -0.1800$,
$h_{2} = 0.0350$, $f_{0} = 2.76$, $f_{1} = 6.79$, $f_{2} = -5.29$, $g_{1} = -0.47$ and
$g_{2} = 1.04$ . This set is used in \cite{hugo14} for the non-conformal holographic plasma.  We also consider this set in our section $IV$.

The pressure is calculated from (\ref{fodortrami}):
\begin{equation}
{p}_{2}(T,\mu)=T^{4}\, \bigintss_{0}^{T} \, dT^{'} \,
{\frac{e^{-h_{1}/{\tau'}-h_{2}/{{\tau'}^{2}}}}{T^{'}}} \cdot \Bigg[h_{0}+
{\frac{f_{0} \cdot \Big[tanh(f_{1}\cdot {\tau'}+f_{2})+1\Big]}{1+g_{1} \cdot {\tau'}+g_{2} \cdot {{\tau'}^{2}}}}   \Bigg]
+{\frac{\chi_{2}}{2}}\, \mu^{2}  T^{2}
\label{prelatt}
\end{equation}
Inserting (\ref{prelatt}) into (\ref{fodortrami}) we find the following expression for the energy density \cite{we2015}:
$$
{\varepsilon}_{2}(T,\mu)=T^{4} \,e^{-h_{1}/{\tau}-h_{2}/{\tau}^{\,2}} \cdot \Bigg[h_{0}+
{\frac{f_{0} \cdot \Big[tanh(f_{1}\cdot {\tau}+f_{2})+1\Big]}{1+g_{1} \cdot {\tau}+g_{2} \cdot {\tau}^{2}}} \Bigg]
+{\frac{\mu^{2}}{2}}\, T^{3} \, {\frac{d \chi_{2}}{d T}}
$$
\begin{equation}
+3\,T^{4}\, \bigintss_{0}^{T} \, dT^{'} \,
{\frac{e^{-h_{1}/{\tau'}-h_{2}/{{\tau'}^{2}}}}{T^{'}}} \cdot \Bigg[h_{0}+
{\frac{f_{0} \cdot \Big[tanh(f_{1}\cdot {\tau'}+f_{2})+1\Big]}{1+g_{1} \cdot {\tau'}+g_{2} \cdot {{\tau'}^{2}}}}\Bigg]
+{\frac{3 \,\chi_{2}}{2 }} \, \mu^{2} T^{2}
\label{epslatt}
\end{equation}

\begin{figure}[ht!]
\begin{center}
\subfigure[ ]{\label{fig:first7}
\includegraphics[width=0.65\textwidth]{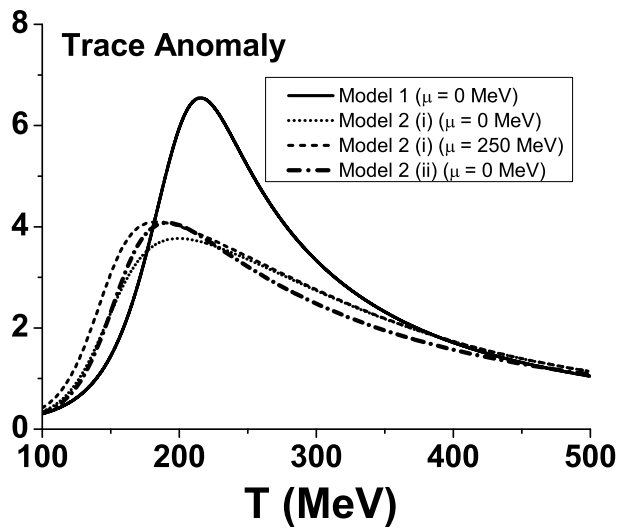}}\\
\end{center}
\caption{Trace anomaly for several chemical potentials.}
\label{fig7}
\end{figure}

In Fig. \ref{fig7} we present the trace anomaly given by $(\varepsilon-3p)/T^{4}$ for {\it{Models}} $1$ and $2$ at zero chemical potential and for Model $2$ at finite chemical potential.

\subsection{Numerical Results }

We start our numerical study following the calculations developed in \cite{raja}, with no bulk effects and considering that the evolution
starts at $\tau_{0}=0.5 \, fm$, with $\Phi(\tau_{0})=0$.  We want to compare different equations of state and determine the effects of certain features of
these EOS (such as, for example, finite chemical potential) on the evolution of the fluid.  To do this we would like to evolve them starting from the same initial
state. However this is not possible because if they represent systems at the same initial temperature, then, these systems will have different initial energy
densities and vice-versa.  In relativistic heavy ion collisions, before thermal equilibrium formation, energy is released from the projectiles in a certain volume
in the central rapidity region. It is perhaps more realistic to say that the quark-gluon system has first a defined energy density determined from the conditions
of the collisions and then it reaches thermal equilibrium, forming a QGP with the properties determined by the equation of state.  Since it is very difficult to
know what comes first, energy density or temperature,  we will test the two possibilities: same initial temperature, $T_{initial}=305 \, MeV$ and same initial
energy density, $\varepsilon_{initial}=16 \, GeV/fm^{3}$.  We will find out that our conclusions regarding cavitation do not depend on  this choice and are the
same for different initial conditions.

\begin{figure}[ht!]
\begin{center}
\subfigure[ ]{\label{fig:first1}
\includegraphics[width=0.65\textwidth]{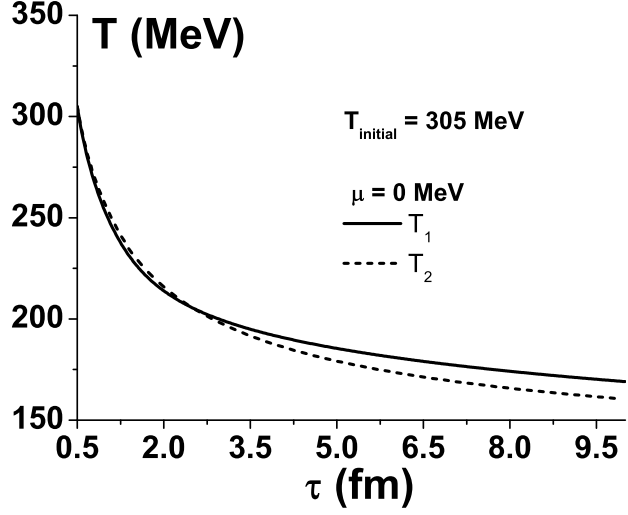}}\\
\subfigure[ ]{\label{fig:second1}
\includegraphics[width=0.65\textwidth]{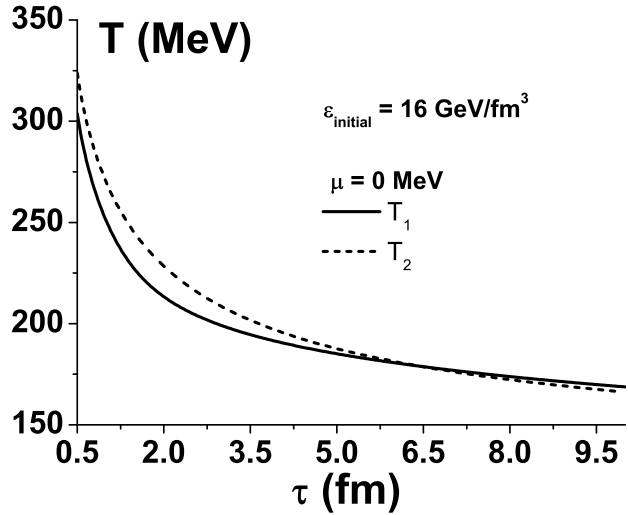}}
\end{center}
\caption{Temperature evolution with no bulk effects for {\it{Model 1}} and {\it{Model 2 (i)}} from $\tau_{0}=0.5 \, fm$ to $\tau_{final}=10 \, fm$ at
zero chemical potential.  a) Same initial temperature.
b) Same initial energy density.}
\label{fig1}
\end{figure}

\begin{figure}[ht!]
\begin{center}
\subfigure[ ]{\label{fig:first2}
\includegraphics[width=0.65\textwidth]{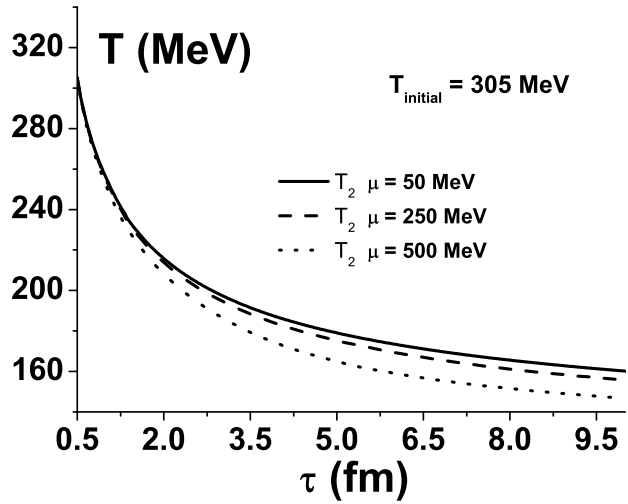}}\\
\subfigure[ ]{\label{fig:second2}
\includegraphics[width=0.65\textwidth]{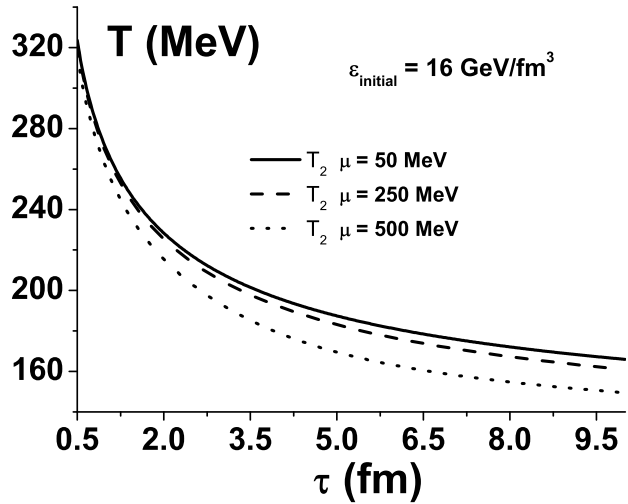}}
\end{center}
\caption{Temperature evolution with no bulk effects for {\it{Model 2 (i)}} from $\tau_{0}=0.5 \, fm$ to $\tau_{final}=10 \, fm$
at finite chemical potential.  a) Same initial temperature.
b) Same initial energy density. }
\label{fig2}
\end{figure}

We solve the coupled  equations (\ref{evoener})
and (\ref{evophi}) to obtain the temperature for each model considered.  The temperature evolution at zero chemical potential is showed in
Fig. \ref{fig1}.
The effects of the chemical potential  are presented in Fig. \ref{fig2} for {\it{Model 2}}. From the figure we clearly see that systems with larger chemical potential cool faster.

Cavitation is an effect mostly associated with peaks of the bulk viscosity close to phase transition \cite{raja,romacav,deni15}.
We now consider bulk effects
via eq. (\ref{evopi}) with the relaxation scale approximated by $\tau^{\eta}_{\Pi} = \tau^{\zeta}_{\Pi}$.  As previously mentioned,
cavitation is generated when the
longitudinal pressure, $P_{\xi}$, is negative and it starts when $P_{\xi}(\tau_{cav})=0$ in (\ref{plon}),  where $\tau_{cav}$ is the
time at which the longitudinal pressure
vanishes.  In this subsection we will make use of (\ref{etapers}) to (\ref{entd}) to solve numerically the three coupled evolution
equations (\ref{evoener}), (\ref{evophi})
and (\ref{evopi}) and determine the time evolution of the longitudinal pressure (\ref{plon}).

\begin{figure}[ht!]
\begin{center}
\subfigure[ ]{\label{fig:first3}
\includegraphics[width=0.65\textwidth]{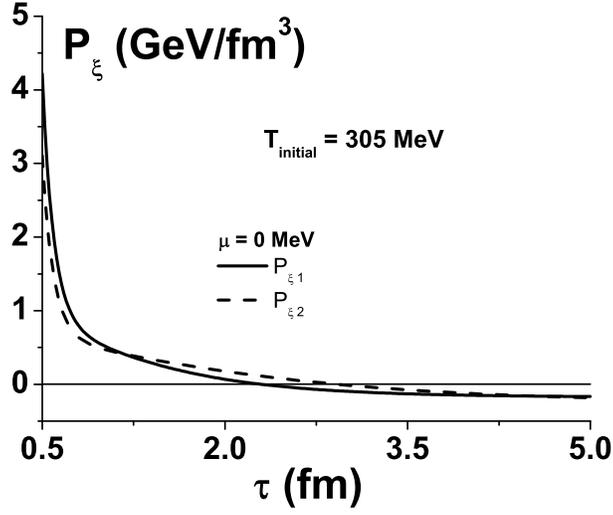}}\\
\subfigure[ ]{\label{fig:second3}
\includegraphics[width=0.65\textwidth]{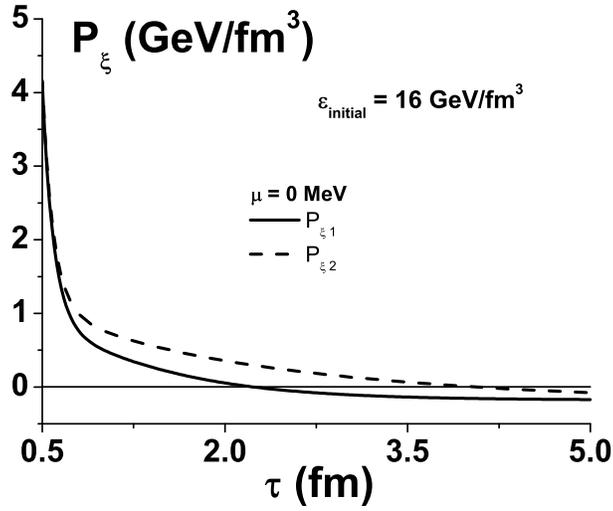}}
\end{center}
\caption{Time evolution of the longitudinal pressure (\ref{plon}) from $\tau_{0}=0.5 \, fm$ and zero chemical potential.
See Tables \ref{tablefig3a} and \ref{tablefig3b}. a) Same initial temperature.
b) Same initial energy density. }
\label{fig3}
\end{figure}

We start with zero chemical potential in Fig. \ref{fig3} for the two models with the initial conditions $\tau_{0}=0.5 \, fm$,
$\Phi(\tau_{0})=0$, $\Pi(\tau_{0})=0$ for
$T_{initial}=305 \, MeV$ and also for $\varepsilon_{initial}=16 \, GeV/fm^{3}$.

\begin{table}[!htbp]
\caption{Time when cavitation starts  (in Fig. \ref{fig:first3} )}
\vspace{0.3cm}
\centering
\begin{tabular}{cc}
\hline
\hspace{1.0cm}  &  $\tau_{cav}$ $[fm]$ \, (fixed $T_{initial}=305 \, MeV$)  \hspace{1.0cm}  \\ [0.8ex]
\hline
\hline
{\it{Model 1}} &  \hspace{1.0cm} 2.3   \\
\hline
{\it{Model 2 (i)}} & \hspace{1.0cm}  2.9   \\
\hline
\end{tabular}
\label{tablefig3a}
\end{table}

\begin{table}[!htbp]
\caption{Time when cavitation starts  (in Fig. \ref{fig:second3} )}
\vspace{0.3cm}
\centering
\begin{tabular}{cc}
\hline
\hspace{1.0cm}  &  $\tau_{cav}$ $[fm]$ \, (fixed $\varepsilon_{initial}=16 \, GeV/fm^{3}$)  \hspace{1.0cm}  \\ [0.8ex]
\hline
\hline
{\it{Model 1}} &  \hspace{1.0cm} 2.22   \\
\hline
{\it{Model 2 (i)}} & \hspace{1.0cm}  4.06   \\
\hline
\end{tabular}
\label{tablefig3b}
\end{table}

\begin{figure}[ht!]
\begin{center}
\subfigure[ ]{\label{fig:first4}
\includegraphics[width=0.65\textwidth]{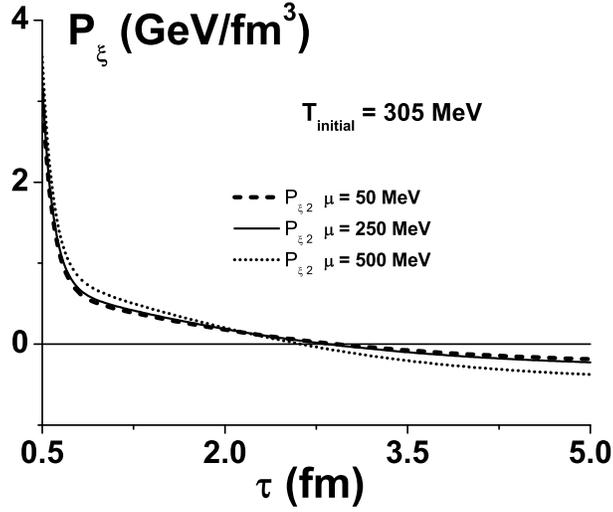}}\\
\subfigure[ ]{\label{fig:second4}
\includegraphics[width=0.65\textwidth]{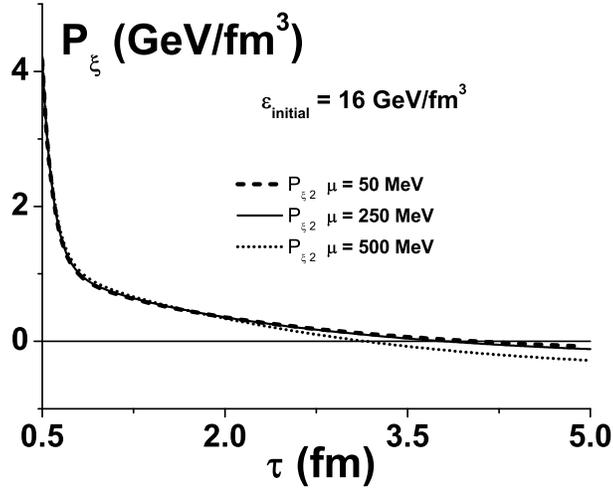}}
\end{center}
\caption{Time evolution of the longitudinal pressure (\ref{plon}) from $\tau_{0}=0.5 \, fm$ and finite chemical potential for the {\it{Model 2}}.
See Tables \ref{tablefig4a} and \ref{tablefig4b}
for values of the chemical potential. a) Same initial temperature.
b) Same initial energy density.}
\label{fig4}
\end{figure}

In Fig. \ref{fig4} we show results at finite chemical potentials for the {\it{Model 2}}. The cavitation starts earlier for increasing the chemical potential.

\begin{table}[!htbp]
\caption{Time when cavitation starts  (in Fig. \ref{fig:first4} ) }
\vspace{0.3cm}
\centering
\begin{tabular}{cc}
\hline
\hspace{1.0cm}  &  $\tau_{cav}$ $[fm]$ \, (fixed $T_{initial}=305 \, MeV$)   \\ [0.8ex]
\hline
\hline
{\it{Model 2 (i)}} \,$(\mu=50 \, MeV)$ &  \hspace{0.8cm} 2.91  \\
\hline
{\it{Model 2 (i)}} \,$(\mu=250 \, MeV)$ & \hspace{0.8cm} 2.84  \\
\hline
{\it{Model 2 (i)}} \,$(\mu=500 \, MeV)$ & \hspace{0.8cm} 2.63  \\
\hline
\end{tabular}
\label{tablefig4a}
\end{table}

\begin{table}[!htbp]
\caption{Time when cavitation starts  (in Fig. \ref{fig:second4} ) }
\vspace{0.3cm}
\centering
\begin{tabular}{cc}
\hline
\hspace{1.0cm}  &  $\tau_{cav}$ $[fm]$ \, (fixed $\varepsilon_{initial}=16 \, GeV/fm^{3}$)   \\ [0.8ex]
\hline
\hline
{\it{Model 2 (i)}} \,$(\mu=50 \, MeV)$ &  \hspace{0.8cm} 4.04  \\
\hline
{\it{Model 2 (i)}} \,$(\mu=250 \, MeV)$ & \hspace{0.8cm} 3.8  \\
\hline
{\it{Model 2 (i)}} \,$(\mu=500 \, MeV)$ & \hspace{0.8cm} 3.15  \\
\hline
\end{tabular}
\label{tablefig4b}
\end{table}

\section{Cavitation in a non-conformal holographic plasma}

One interesting question is whether it is possible to find cavitation within different approaches for the underlying many-body microscopic
dynamics. From the strongly coupled
nature of the QGP, in particular near the deconfinement transition temperature, the microscopic physics should be inherently non-perturbative.
In this context, holographic
methods may shed light upon the
nature of transport coefficients for such strongly coupled plasmas, since these coefficients are difficult to obtain from lattice methods.

We follow the parametrization and second-order theory from \cite{hugo14}. Considering Bjorken symmetry and flat spacetime, our analysis reduces
to solving the following dissipative
equations:
\begin{equation}
\tau_\Pi^{\eta} \left(D\pi^{\langle \mu\nu\rangle}+\frac{4\theta}{3}\pi^{\mu\nu}\right) +\pi^{\mu\nu} = -2\eta \sigma^{\mu\nu}+ \frac{\lambda_1}{\eta^2}
\pi_\lambda^{\langle \mu} \pi^{\nu\rangle\lambda} + \tau_\pi^{*}\pi^{\mu\nu}\,\frac{\Pi}{3\zeta}\,
\label{definenewshearstress}
\end{equation}
and
\begin{equation}
\tau_\Pi^{\zeta} \left(D\Pi+\Pi \theta\right)+\Pi  = -\zeta \theta  +\frac{\xi_1}{\eta^2} \pi_{\mu\nu}\pi^{\mu\nu} +\frac{\xi_2}{\zeta^2}\Pi^2
+\tau_\Pi^{\zeta}
\,\Pi\,\,D \ln \left(\frac{\zeta}{s}\right) \, .
\label{definenewbulk}
\end{equation}
The employed holographic method is a bottom-up solution \cite{gubser-nellore} that parametrizes some of the thermodynamical properties of
lattice (2+1)-flavor QCD near the
crossover phase transition, described by {\it{Model 2 (ii)}} at zero chemical potential with the parameters \cite{ratti,hugo14}.
The trace anomaly with these parameters is shown in Fig. \ref{fig7}.

It is known that holographic calculations of transport coefficients, such as the holographic value
of $\eta/s = 1/4 \pi$ \cite{kss},
are generally consistent with experimental investigations of the evolution of QGP. The  calculations employing  usual kinetic theory result
in larger values of $\eta/s$, which
corroborates that holographic calculations are at least consistent with the QGP evolution near the deconfinement region  \cite{depack}.

The advantage of this approach is that the non-trivial temperature dependence of the transport coefficients were calculated for a phenomenological
bottom-up model and parametrized in \cite{hugo14}. In the next subsection we summarize the relevant results.

\subsection{Transport coefficients and Bjorken flow}

The coefficient $\eta/s$ is still $1/4 \pi$, as is expected for several holographic models \cite{kss, Buchel:2003tz}. The shear relaxation
time is parametrized as follows ($T_c^{h}=143.8$ MeV):
\begin{equation}
\label{eq:kappafit}
\frac{\tau_{\Pi}^{\eta} \, \eta}{T^2}\left(x=\frac{T}{T_c^{h}} \right) = \frac{a_{\eta}^{h}}{1+ e^{b_{\eta}^{h}\left(c_{\eta}^{h}-x\right)} +
e^{d_{\eta}^{h}\left(e_{\eta}^{h}-x\right)} + e^{f_{\eta}^{h}\left(g_{\eta}^{h}-x\right)}},
\end{equation}
where $a_{\eta}^{h}$ to $g_{\eta}^{h}$ are fit parameters listed in Tab. \ref{tab:parametersholo}.

The parametrization of bulk viscosity is:
\begin{equation}
\label{eq:zetafit}
\frac{\zeta}{s}\left(x=\frac{T}{T_c^{h}}\right) = \frac{a_{bulk}^{h}}{\sqrt{\left(x-b_{bulk}^{h}\right)^2+{(c_{bulk}^{h})}^2}} +
\frac{d_{bulk}^{h}}{x^2+{(e_{bulk}^{h})}^2},
\end{equation}
where $a_{bulk}^{h}$ to $e_{bulk}^{h}$ are fit parameters in Tab. \ref{tab:parametersholo}.

In order to parametrize the bulk relaxation time, we adopt a causality bound, relating it to the shear relaxation time \cite{hugo14}:
\begin{equation}
\label{eq:tauPIfit}
\tau_{\Pi}^{\zeta} T \left(x=\frac{T}{T_c^{h}}\right) = \frac{a_{\zeta}^{h}}{\sqrt{\left(x-b_{\zeta}^{h}\right)^2+({c_{\zeta}^{h})}^{\,2}}} +
\frac{d_{\zeta}^{h}}{x},
\end{equation}
with the corresponding fit parameters $a_{\zeta}^{h}$ to $d_{\zeta}^{h}$ being given in Tab. \ref{tab:parametersholo}.

\begin{table}[htp]
\caption{Parameters: $\tau_{\Pi}^{\eta} \, \eta/T^2$ for Eq.\ \eqref{eq:kappafit}, \\ $\zeta/s$ for Eq.\ \eqref{eq:zetafit}\, and \,
$\tau_{\Pi}^{\zeta}T$ for Eq.\ \eqref{eq:tauPIfit}. }
\begin{center}
\begin{tabular}{cccccccc}
\hline
\hline
$a_{\eta}^{h}$ & $b_{\eta}^{h}$ & $c_{\eta}^{h}$ & $d_{\eta}^{h}$ & $e_{\eta}^{h}$ & $f_{\eta}^{h}$  & $g_{\eta}^{h}$ \\ \hline
0.2664 \, & 2.029 \, & 0.7413 \, & 0.1717 \, & -10.76 \, & 9.763 \, & 1.074 \\
\hline\hline\hline
$a_{bulk}^{h}$ & $b_{bulk}^{h}$ & $c_{bulk}^{h}$ & $d_{bulk}^{h}$ & $e_{bulk}^{h}$ & &   \\ \hline
0.01162 \, & 1.104 \, & 0.2387 \, & -0.1081 \, & 4.870 & &  \\
\hline\hline\hline
$a_{\zeta}^{h}$ & $b_{\zeta}^{h}$ & $c_{\zeta}^{h}$ & $d_{\zeta}^{h}$ & & & \\ \hline
0.05298 \, & 1.131 \, & 0.3958  \, & -0.05060 & & & \\ \hline\hline
\end{tabular}
\end{center}
\label{tab:parametersholo}
\end{table}

The other coefficients are determined in a phenomenological approach. However there is still a non-trivial behavior near the transition temperature.
These are estimates considering fluid-gravity calculations for conformal fluid, as well as other strongly coupled holographic calculations
\cite{Bhattacharyya:2008jc, brsss, Kanitscheider:2009as}. We consider then the following \cite{hugo14}:
\begin{eqnarray}
\lambda_{1} = \frac{2 \eta^{2}}{s T} \, , \nonumber \\
\tau_{\pi}^{*} = - 3 \tau_{\Pi}^{\eta} \left( \frac{1}{3} - c_{s}^{2} \right) \, , \nonumber \\
\xi_{1} = \lambda_{1} \left( \frac{1}{3} - c_{s}^{2} \right) \, , \nonumber \\
\xi_{2} = 2 \eta \tau_{\Pi}^{\zeta} \,c_{s}^{2} \left( \frac{1}{3} - c_{s}^{2} \right)  \, . \nonumber \\
\end{eqnarray}

\begin{figure}[ht!]
\begin{center}
\subfigure[ ]{\label{fig:first6}
\includegraphics[width=0.65\textwidth]{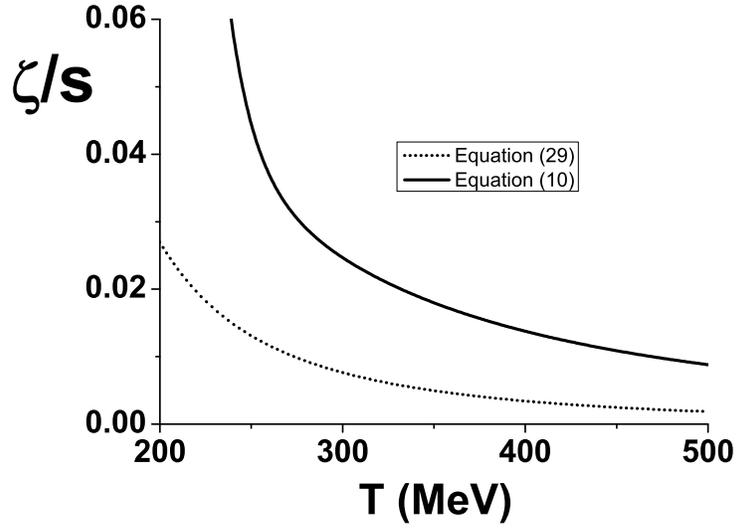}}\\
\subfigure[ ]{\label{fig:second6}
\includegraphics[width=0.65\textwidth]{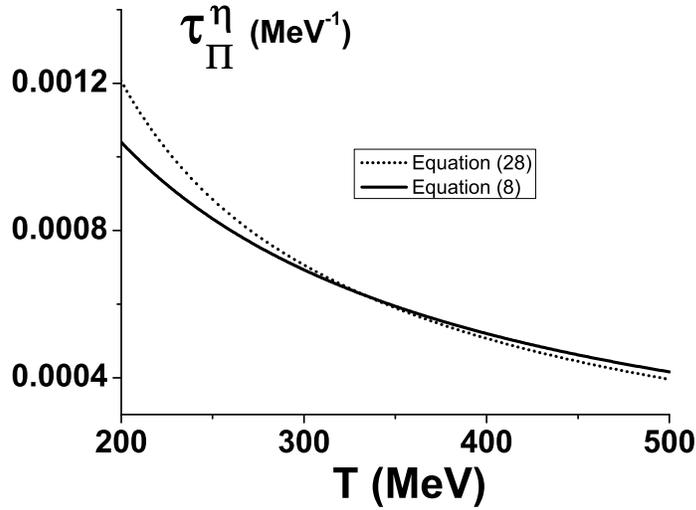}}
\end{center}
\caption{(a) Bulk per entropy density. (b) Relaxation scale.  The parametrizations come from
\cite{raja,hugo14,deni15}.}
\label{fig6}
\end{figure}

In Fig. \ref{fig:first6} we show the two bulk expressions given by (\ref{zetapers}) and (\ref{eq:zetafit}) using the parametrizations from \cite{raja,hugo14,deni15}.  Notice that near the deconfinement
temperature the bulk viscosity per entropy density of the holographic calculation given by (\ref{eq:zetafit}) is significantly smaller than (\ref{zetapers}).
Both come from distinct microscopic approaches  and the difference between them reflects our lack of knowledge on the subject.
We hope that future theoretical  and experimental work can shed light upon bulk viscosity properties of strongly coupled plasmas. Similarly, in Fig. \ref{fig:second6} we present the two relaxation times as a function of the temperature, (\ref{taueta}) and (\ref{eq:kappafit}). Eq.(\ref{taueta}) is the result of a conformal calculation and  Eq. (\ref{eq:kappafit}) includes non-conformal contributions calculate through holography.

It is straightforward to solve Eqs. (\ref{definenewshearstress}) and (\ref{definenewbulk}) in the Bjorken flow. We follow the conventions
adopted in the previous section, so the evolution equations are (\ref{evoener}) and the following:
\begin{equation}
\tau^{\eta}_{\Pi} \frac{\partial \Phi}{\partial \tau} = \frac{4 \eta}{3 \tau} - \Phi - \left[ \frac{4 \tau^{\eta}_{\Pi}}{3 \tau} \Phi +
\frac{\lambda_{1}}{2 \eta^2} \Phi^2 \right] + \frac{\tau_{\pi}^{*}}{3 \zeta} \Phi \Pi
\label{evophia}
\end{equation}
and
\begin{equation}
\tau^{\zeta}_{\Pi} \frac{\partial \Pi}{\partial \tau} +\tau^{\zeta}_{\Pi}{\frac{\Pi}{\tau}}= - \frac{\zeta}{\tau} - \Pi + \frac{3}{2}
\frac{\xi_{1}}{\eta^{2}} \Phi^{2} + \frac{\xi_{2}}{\zeta^{2}} \Pi^{2}
+ \frac{\tau_{\Pi}^{\zeta}}{(\zeta/s)} \Bigg[{\frac{\partial}{\partial T}} \left( \frac{\zeta}{s} \right) \Bigg] {\frac{\partial T}{\partial \tau}} \, .
\label{evopia}
\end{equation}
In particular, equations (\ref{evophia}) and (\ref{evopia}) are the same as (\ref{evophi}) and (\ref{evopi}) but with extra transport coefficients.
One interesting feature of such
second order theory is that both bulk and shear evolution equations are nonlinear and coupled to each other.  The entropy density in (\ref{eq:zetafit}),
(\ref{evophia}) and (\ref{evopia}) is calculated via (\ref{entd}) using the equation of state of the Model $2$.

We solve numerically Eqs. (\ref{evoener}), (\ref{evophia}) and (\ref{evopia}) in the Bjorken flow and with the discussed holographic transport
coefficients to search possible
indications of cavitation.  We consider the initial conditions $\tau_{0}=0.5 \, fm$, $\Phi(\tau_{0})=0$, $\Pi(\tau_{0})=0$ for both initial approaches
$T_{initial}=305 \, MeV$ and also $\varepsilon_{initial}=16 \, GeV/fm^{3}$.

\begin{figure}[ht!]
\begin{center}
\subfigure[ ]{\label{fig:first5}
\includegraphics[width=0.65\textwidth]{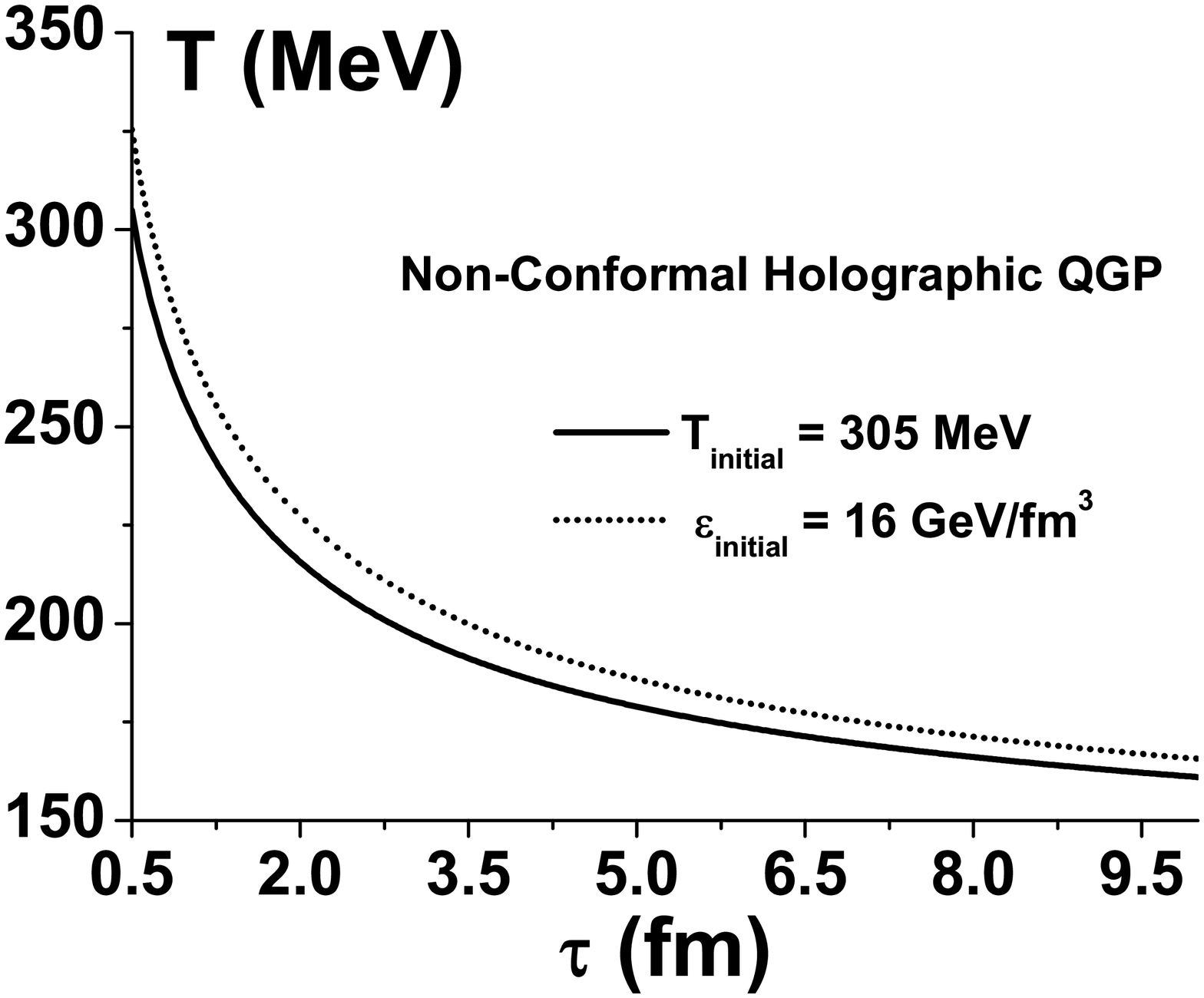}}\\
\subfigure[ ]{\label{fig:second5}
\includegraphics[width=0.65\textwidth]{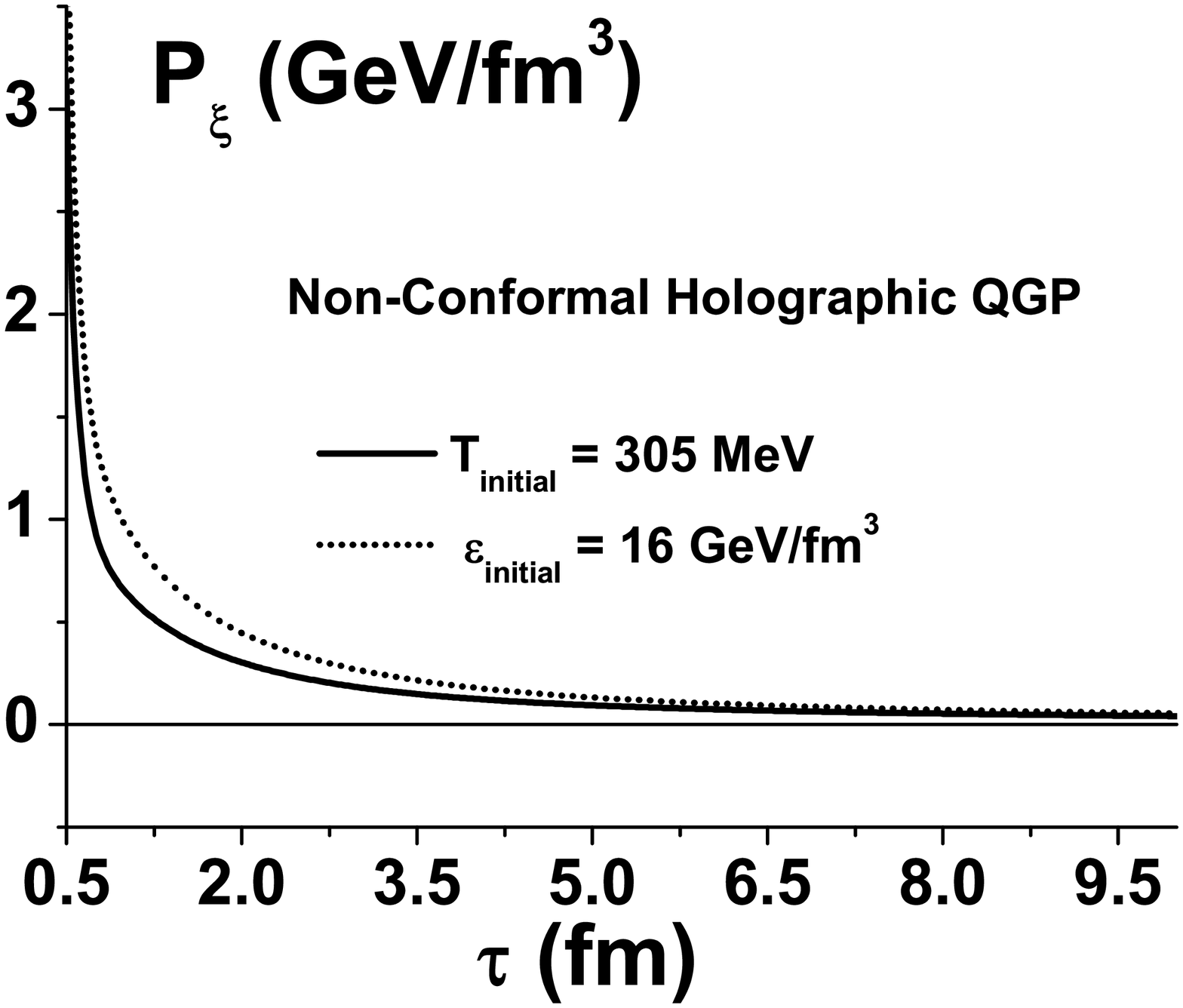}}
\end{center}
\caption{(a) Temperature evolution for the non-conformal holographic QGP. (b)
Time evolution of the longitudinal pressure (\ref{plon}) for the non-conformal holographic QGP at zero chemical potential.}
\label{fig5}
\end{figure}

As can be seen in Fig. \ref{fig5} cavitation does not occur. It is often the case with holographic transport coefficients that dissipative
effects seem to be smaller
than others obtained from different methods, such as kinetic theory calculations. Nonetheless, the holographic values for small ratio of shear
viscosity per entropy density
are consistent so far with experimental results, and thus we suggest that, if confirmed by experimental data, the absence of cavitation may be
a consequence of the strongly coupled  nature of QGP.

\begin{figure}[ht!]
\begin{center}
\subfigure[ ]{\label{fig:first8}
\includegraphics[width=0.65\textwidth]{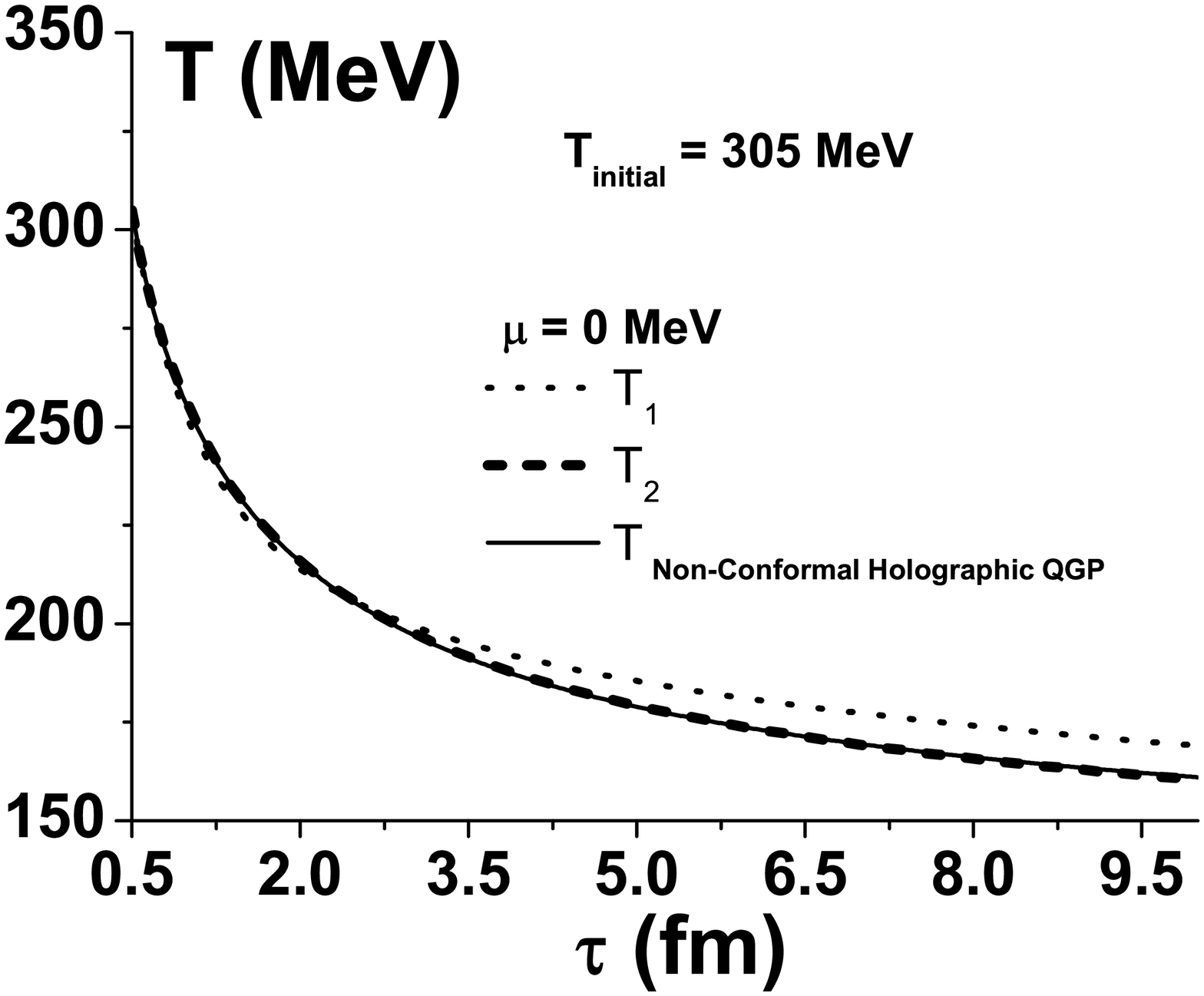}}\\
\subfigure[ ]{\label{fig:second8}
\includegraphics[width=0.65\textwidth]{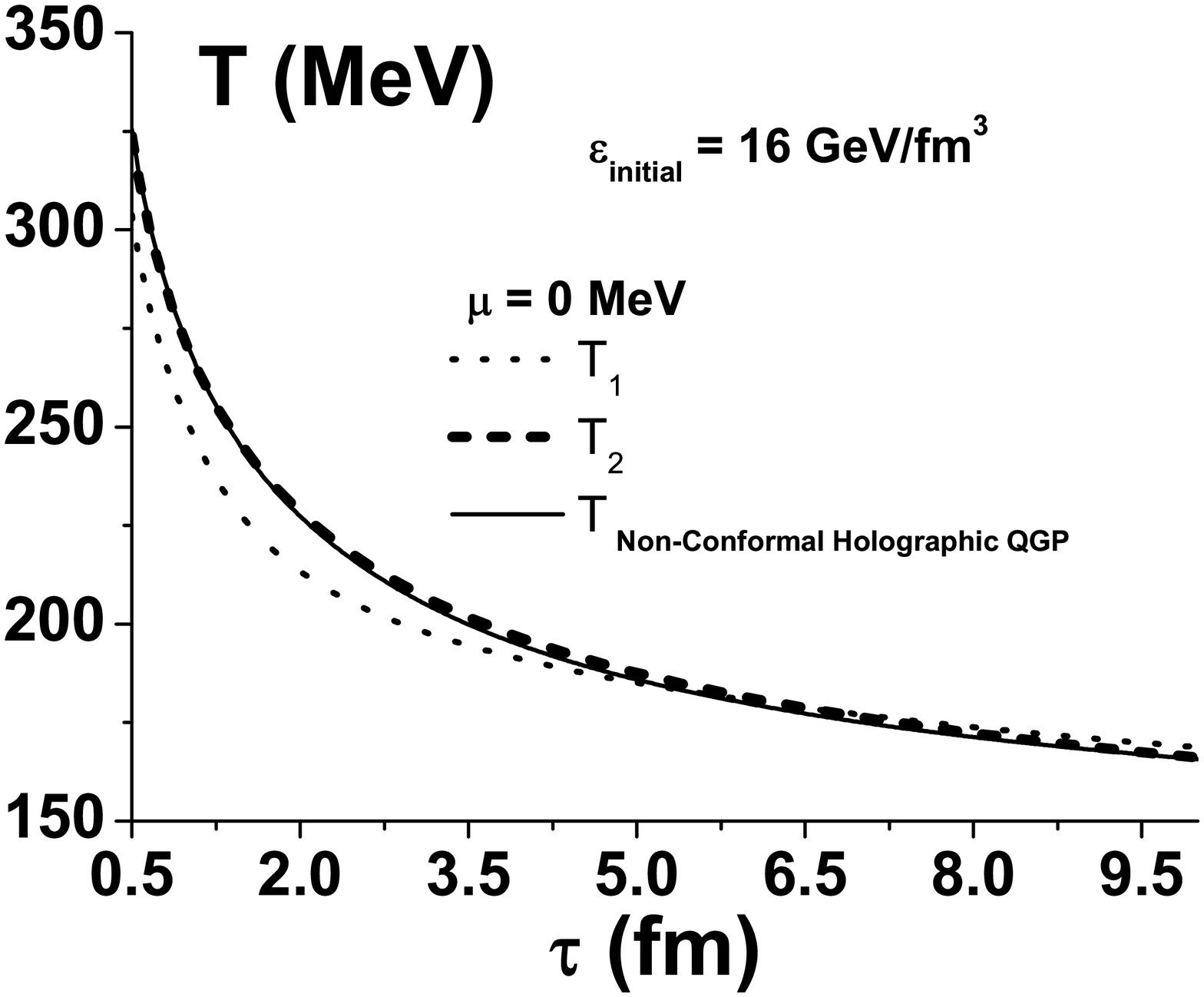}}
\end{center}
\caption{Temperature evolution with (a) same initial temperature and (b) same initial energy density.}
\label{fig8}
\end{figure}

\begin{figure}[ht!]
\begin{center}
\subfigure[ ]{\label{fig:first9}
\includegraphics[width=0.65\textwidth]{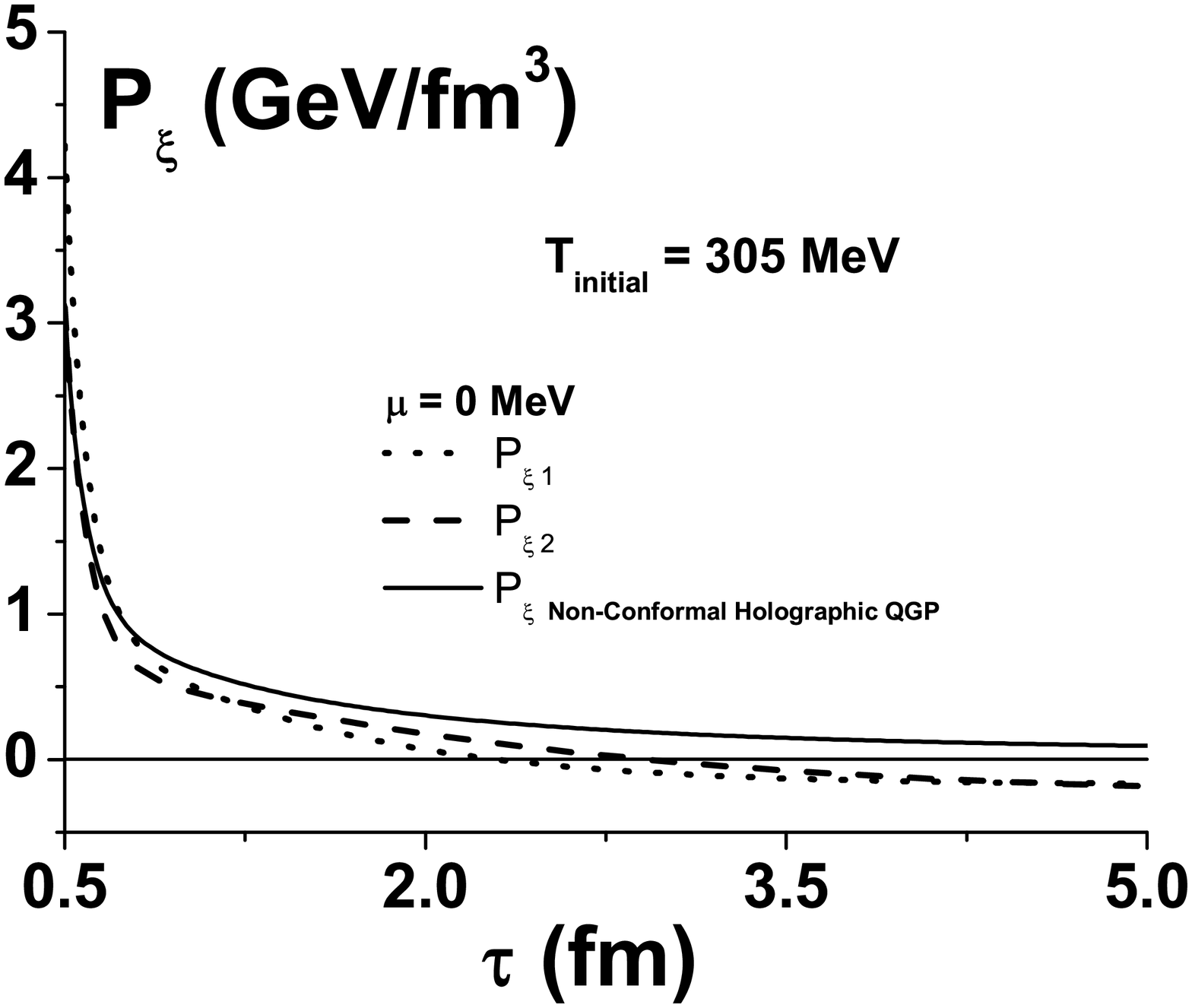}}
\subfigure[ ]{\label{fig:second9}
\includegraphics[width=0.65\textwidth]{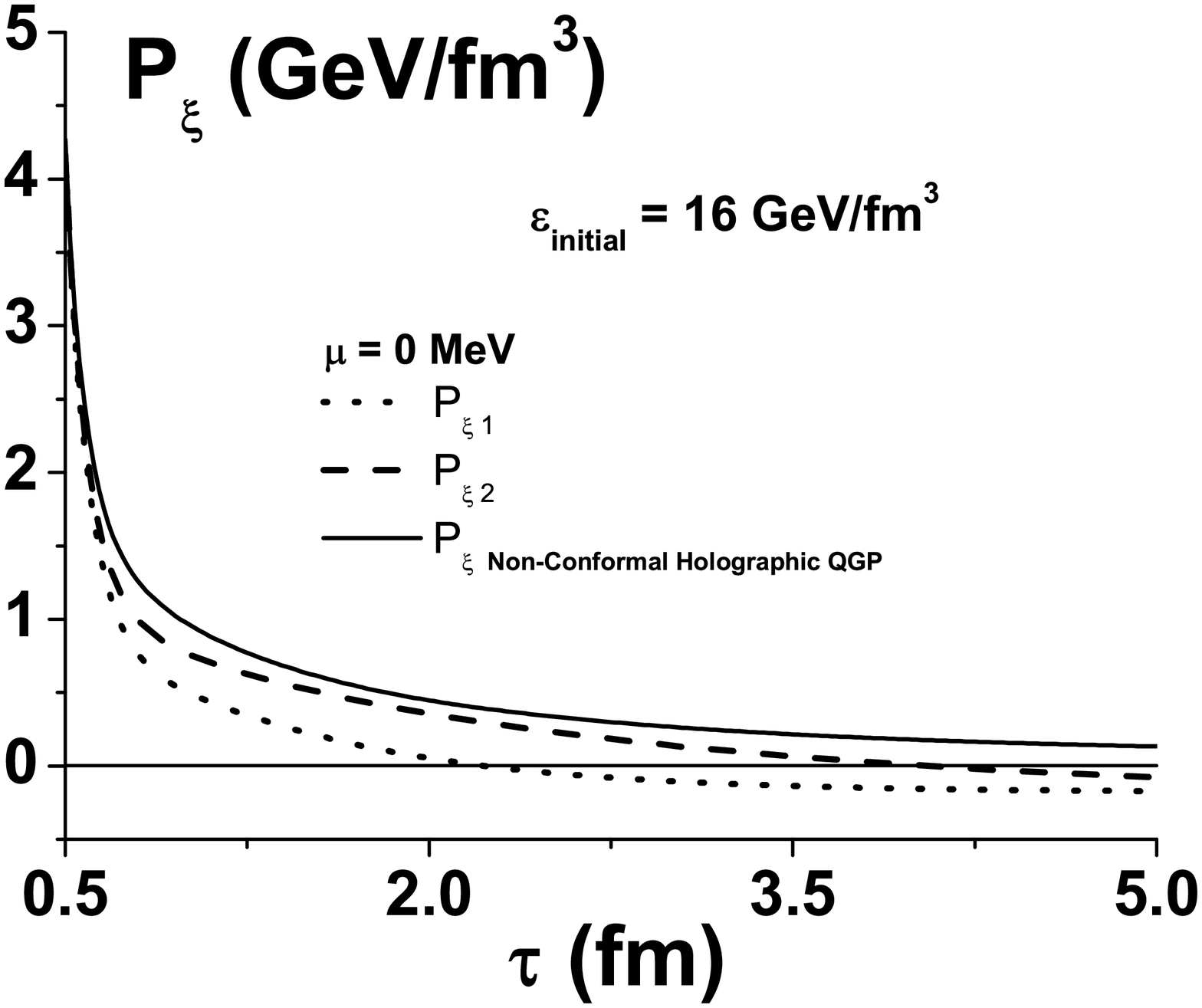}}
\end{center}
\caption{Longitudinal pressure evolution with  (a) same initial temperature and (b) same initial energy density.}
\label{fig9}
\end{figure}

In order to compare the Models $1$ and $2$ with the non-conformal holographic Model we show again, in Fig. \ref{fig8} and in Fig. \ref{fig9} the same plots of the temperature and longitudinal pressure evolution
previously presented in Fig. \ref{fig1}, Fig. \ref{fig3} and Fig. \ref{fig5}.

\section{Conclusions}

In this work, we considered several equations of state at different chemical potentials in order to investigate the occurrence of cavitation
in QGP. This was done through a numerical integration of the Bjorken hydrodynamical equations and determination of the pressure as a function of
proper time.

We used a phenomenological parametrization of the lattice QCD equation of state with two sets of transport coefficients. For the  bulk viscosity
we used a parametrization of results obtained in a Monte-Carlo simulation of pure gluon dynamics SU(3) \cite{meyer07,raja}.
For both lattice QCD and mean field QCD equations of state, cavitation occurs at smaller time scales as the chemical potential increases.

As is often the case with transport coefficients calculated with
lattice methods near the  phase transition,
the height of the bulk per entropy density peak is not generally well established,  most calculations rely on pure glue dynamics and consequently on the
existence of first order phase transitions. We considered a phenomenological holographic calculation of transport coefficients of a non-conformal
QGP \cite{hugo14}.
In our approach, we conclude that cavitation does not occur in the holographic prescription, specially due to the small bulk viscosity as shown in Fig. \ref{fig:first6} and by Eq.  (\ref{eq:zetafit}).  Even with the
addition of more transport coefficients (which
contain peaks near the crossover phase transition) this approach does not support cavitation.
In the future it may be possible to compare these distinct cavitation
results in the scenario where lattice
calculations of bulk effects can consistently consider the crossover phase transition.
We hope that once the magnitude of bulk effects in the QGP is more precisely calculated, we can be more conclusive on whether cavitation occurs.

It should be interesting to generalize our calculations to more complex flows and check whether cavitation can be obtained in different regimes.
Also, a non trivial check of the observable effects of cavitation in QGP would be the study of stability properties of hadron gas bubbles, as well as how
their collapse influences the overall evolution of the plasma. We leave such investigations for future work.

\begin{acknowledgments}
This work was partially financed by the Brazilian funding agencies CAPES, CNPq and FAPESP. We thank J. Noronha and G. Torrieri for instructive discussions.
\end{acknowledgments}

\end{document}